\documentclass[pre,reprint,twocolumn,superscriptaddress,aps,showkeys]{revtex4-2}

\usepackage{graphicx}
\usepackage{amsmath}
\usepackage{amssymb}
\usepackage{color}
\usepackage{siunitx}
\usepackage{nicefrac}
\usepackage[capitalize]{cleveref}

\newcommand{\vb}[1]{\boldsymbol{#1}}
\renewcommand{\d}{\text{d}}
\newcommand{\dbar}{{d\mkern-7mu\mathchar'26\mkern-2mu}}

\begin{document}


\title{Magnetically Driven Elastic Microswimmers: Exploiting Hysteretic Collapse for Autonomous Propulsion and Independent Control.}

\author{Theo Lequy}
\email{tlequy@ethz.ch}
\affiliation{Eidgen{\"o}ssische Technische Hochschule Zürich, R{\"a}mistrasse 101, 8092 Z\"urich, Switzerland }

\author{Andreas M. Menzel}
\email{a.menzel@ovgu.de}
\affiliation{Institut f{\"u}r Physik, 
Otto-von-Guericke-Universit\"at Magdeburg, Universit\"atsplatz 2, 39106 Magdeburg, Germany}

\date{\today}

\begin{abstract}
\textbf{Abstract:} When swimming at low Reynolds numbers, inertial effects are negligible and reciprocal movements cannot induce net motion. 
Instead, symmetry breaking is necessary to achieve net propulsion. Directed swimming can be supported by magnetic fields, which simultaneously provide a versatile means of remote actuation. 
Thus, we analyze the motion of a straight microswimmer composed of three magnetizable beads connected by two elastic links. The swimming mechanism is based on oriented external magnetic fields that oscillate in magnitude. Through induced reversible hysteretic collapse of the two segments of the swimmer, the two pairs of beads jump into contact and separate nonreciprocally. 
Due to higher-order hydrodynamic interactions, net displacement results after each cycle. Different microswimmers can be tuned to different driving amplitudes and frequencies, allowing for simultaneous independent control by just one external magnetic field. 
The swimmer geometry and magnetic field shape are optimized for maximum swimming speed using an evolutionary optimization strategy.
Thanks to the simple working principle, an experimental realization of such a microrobot seems feasible and may open new approaches for microinvasive medical interventions such as targeted drug delivery.
\end{abstract}


\maketitle

\section{Introduction}

The field of microrobotics promises significant advances in biomedicine, because it allows microinvasive interventions, such as targeted drug or cell delivery \cite{jang2019targeted, nelson2023delivering}.
Such microscopic robots could further be used for micromanipulation and targeted removal of toxic substances in the environment \cite{shen2023magnetically}. 
Despite the benefits of the small length scales of microrobots on the order of micrometers, these scales also impose restrictions on design and mode of operation.

Here, we mostly address the central challenges of actuation and swimming at low Reynolds numbers along a requested direction. 
As it becomes very difficult to equip such small robots with on-board power devices, and since potential fuel usually cannot be distributed in their environment in medical applications, most microrobots resort to remote actuation. Besides visible light \cite{palagi2019light} and ultrasound \cite{schrage2023ultrasound}, magnetic fields are a popular choice \cite{mag1994manip}. They can induce magnetic moments in magnetizable materials and exert torques by aligning them with the direction of the magnetic field. Moreover, net forces can be achieved through magnetic field gradients. 
Yet, this external actuation, at first glance, poses challenges on the construction of a microrobot. All magnetizable components of the microswimmer (in our case three magnetizable beads) are addressed simultaneously by the external field. So how can we achieve control over the different internal degrees of freedom to achieve required non-simultaneous patterns of motion, although they share the same actuation channel via the magnetic field? 
Moreover, it appears difficult to steer individual microswimmers separately when multiple of them are present in one external magnetic field.

Generally, swimming at low Reynolds numbers differs from our daily experience at macroscopic scales. In Stokes flow at small length scales, viscous forces dominate and fluid inertia is negligible, causing flows to be time-reversible. 
A sequence of shape changes that is identical when played forward and backward, known as reciprocal motion, therefore does not generate time asymmetry, even if performed at different speed. As a result, the net displacement after a full reciprocal cycle is zero \cite{purcell1977life}. 
Effective propulsion in such environments requires non-reciprocal deformation that breaks time-reversal symmetry. One single linear degree of freedom is insufficient, because any cyclical motion is reciprocal in that case.

Designing microswimmers is thus an act of balance between simplicity, to facilitate fabrication and operation, and complexity, to achieve nonreciprocal motion or otherwise break forward--backward symmetry. 
Nature has evolved various strategies to achieve this, including rotating helical flagella, snake-like undulations, or shape morphing \cite{elgeti2015physics}. Some of them can be mimicked in artificial microswimmers \cite{avron2005pushmepullyou, nelson2009abf, box2017motion, dreyfus2005microscopic}.

A minimal model microswimmer that can achieve net propulsion at low Reynolds numbers is the three-sphere swimmer introduced by Najafi and Golestanian \cite{najafi2004simple}. It consists of three collinearly arranged spheres connected by two arms of variable length. 
By changing the lengths of the two arms in a non-reciprocal manner, net motion is achieved. This model has been studied extensively \cite{najafi2004simple, golestanian2008analytic, leoni2009basic, alexander2009hydrodynamics, pande2015forces, babel2016dynamics, daddi2018state, daddi2018swimming}. 

Although this linear arrangement lends itself to analytical treatment and is easy to visualize, its practical realization remains a challenge. Previous experimental attempts have involved optical tweezers to replicate the movement pattern \cite{leoni2009basic}, which is however impractical for in vivo applications. 
The implementation of a self-assembled microswimmer has been achieved relying on capillary forces at fluid interfaces combined with periodic magnetic actuation \cite{grosjean2016realization}. Yet, it seems that involving the role of fluid surfaces may limit applications inside biological systems. Symmetry breaking was achieved by using spheres of different sizes while tuning the actuation frequency accordingly, and the listed model equations involve inertial effects.

In the following, we propose a different approach, still based on magnetic actuation. In our analysis, we connect the three magnetizable spheres using two elastic springs. We entirely concentrate on the overdamped regime, which generally renders nonreciprocality a challenge. 
To succeed, we focus on an effect investigated earlier in theory and experiments. It refers to the reversible collapse and reseparation of pairs of elastically linked beads, which can be induced magnetically \cite{annunziata2013ferrogel, biller2014modeling, biller2015mesoscopic, puljiz2018reversible, goh2018dynamics}. 
The hysteretic nature of this phenomenon allows the collapse and reseparation to occur at different field strengths, thus breaking time-reversal symmetry.

First, we analytically describe the hysteretic behavior of the collapse of two spheres linked by a finitely extensible spring in Sec.~\ref{sec:hysteretic}. 
Then, we introduce the complete three-sphere swimmer and find conditions for nonreciprocal motion, see Sec.~\ref{sec:non-reciprocal}. 
We proceed by analyzing hydrodynamic interactions in Sec.~\ref{sec:hydrodynamics} and provide a geometric illustration for how the loop in the configuration space leads to net propulsion.
Using an evolutionary strategy algorithm, we optimize the design of the microrobot and the driving magnetic field to achieve maximum speed in Sec.~\ref{sec:evolution}. 
Finally, we conclude in Sec.~\ref{sec:conclusion} with brief ideas on potential implementations and improvements in design. 

\section{Results and Discussion}
\subsection{Hysteretic reversible magnetically induced collapse}
\label{sec:hysteretic}

\begin{figure}
    \centering
    \includegraphics[width=1\linewidth]{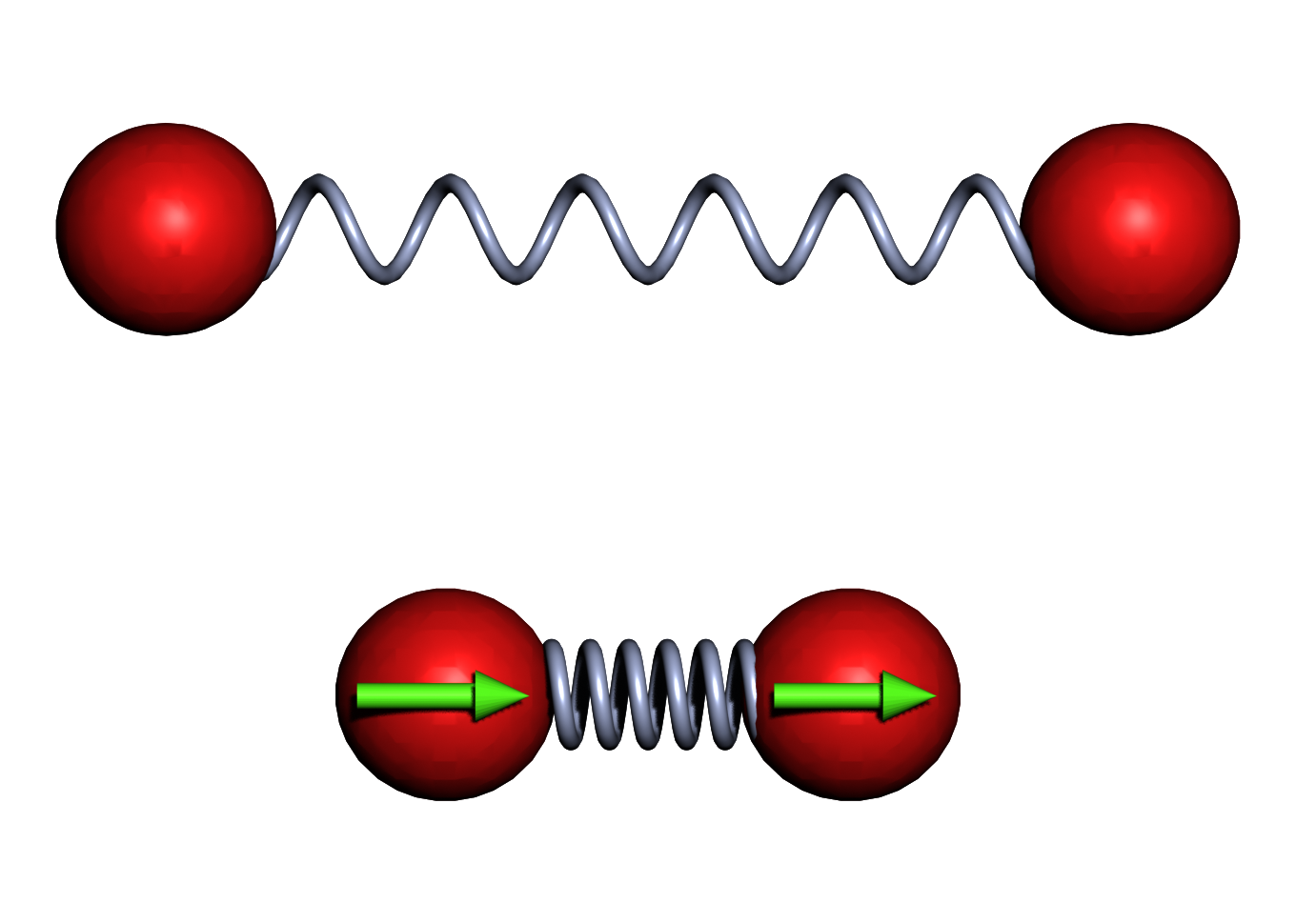}
    \caption{Two magnetic spheres (red) linked by a finitely extensible spring (silver-blue). The top snapshot shows the nonmagnetized configuration with the spring in its undeformed state. In the bottom constellation, a magnetic field is applied along the axis connecting the centers of the two spheres. It induces the magnetic moments $\vb{m}$ (green arrows) in the spheres, leading to an attractive magnetic force, which contracts the spring. Due to the finite extensibility of the spring, the spheres cannot be brought arbitrarily close together.}
    \label{fig:2spheres}
\end{figure}

We start by an analytical consideration of the hysteretic collapse and reseparation of two magnetized spheres linked by a finitely extensible spring shown in \cref{fig:2spheres}. 
The spheres of radius $a$ consist of a superparamagnetic material with high magnetic susceptibility $\chi \gg 1$.  
When applying a homogeneous external magnetic field $\vb{H} = H \vb{\hat{x}}$ in the $x$-direction, each sphere develops a magnetic dipole moment $\vb{m} = \chi_a V \vb{H}$, where $V = 4\pi a^3/3$ is the volume of the sphere. We hereby neglect mutual magnetizing effects among the spheres. That is, we assume that their separations are sufficiently large, and/or the external magnetic field is sufficiently strong to magnetize the spheres towards saturation.
The apparent susceptibility $\chi_a = \chi/(1+N\chi)$ is reduced due to demagnetization, with a demagnetizing factor $N=1/3$ for spheres. Thus, the attractive dipole--dipole interaction between the two spheres can be described by the potential
\begin{equation}
    U_{\text{mag}} = -\frac{8\pi}{9} \chi_a^2 \mu_0 a^6 \frac{H^2}{x^3},
\end{equation}
where $x$ is the center-to-center distance between the spheres and $\mu_0$ is the vacuum permeability.
To counterbalance this attraction, the spheres are connected by a finitely extensible nonlinear elastic spring (FENE) \cite{warner1972kinetic} of rest length $\ell$ in the undeformed state, maximum change in length $\ell_{\text{max}}$ under extension or compression, and stiffness $k$, with the spring potential
\begin{equation}\label{eq:U_FENE}
    U_{\text{FENE}} =
        -\frac{1}{2} k\ell_{\text{max}}^2 \ln\left[1 - \frac{(x - \ell)^2}{\ell_{\text{max}}^2}\right].
\end{equation}

\begin{figure*}
    \centering
    \includegraphics[width=\linewidth]{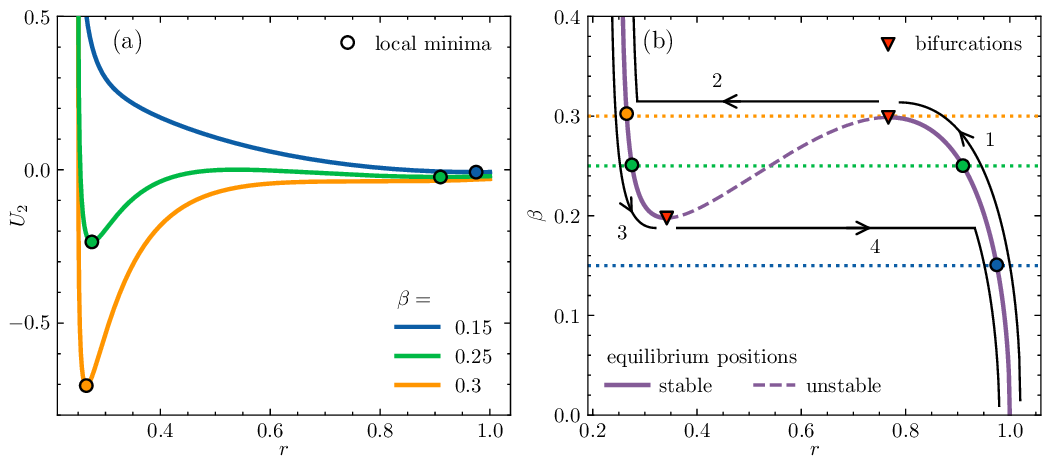}
    \caption{
         (a) Two magnetizable spheres are subject to the dimensionless potential $U_2$, see \cref{eq:2sphere_potential}, when linked by a FENE spring and exposed to an external magnetic field of rescaled strength $\beta$. 
         $r$ denotes the rescaled center-to-center distance between the spheres.
         The extensibility of the spring is set to $\epsilon=0.75$. 
         When the magnetic field strength $\beta$ is varied, different local minima (circles) appear. 
         For $\beta=0.15$ (blue), there is only one stable configuration close to the rest length of the spring, $r\approx1$.
         When increasing $\beta$ to $0.25$ (green), the minimum moves slightly to the left and a second minimum appears in a compressed state of the spring, $r\approx0.25$,
         due to the induced magnetic attraction. 
         For $\beta = 0.3$ (orange), the magnetic attraction dominates, leading to the disappearance of the minimum on the right so that the compressed state is the only stable minimum. 
        (b) Bifurcation diagram for this configuration in the plane spanned by the rescaled length of the spring $r$ and the rescaled magnitude of the external magnetic field $\beta$, see \cref{eq:beta_bif}. 
         1. When the magnetic field is increased, the stable equilibrium position (solid purple line) moves to smaller lengths $r$ of the spring, until the solution vanishes in a saddle-node bifurcation (red triangle). 2. From there, the spheres collapse to a compressed state of the spring.
         3. When the field is decreased again, the spheres remain in the collapsed state for magnetic field amplitudes $\beta$ lower than those of the previous event of collapse, implying hysteresis. 4. Finally, another saddle-node bifurcation appears, and the spheres abruptly detach and reseparate.
         At intermediate field strengths $\beta$, between both saddle-node bifurcations, two stable equilibria coexist, separated by an unstable solution (dashed purple line). 
         The three different field strengths $\beta$ in (a) are marked by horizontal, dotted, colored lines. Corresponding minima in the potential in (a) are indicated by circled, colored dots.
        }
    \label{fig:pot_bif}
\end{figure*}

\begin{figure*}
    \centering
    \includegraphics[width=\linewidth]{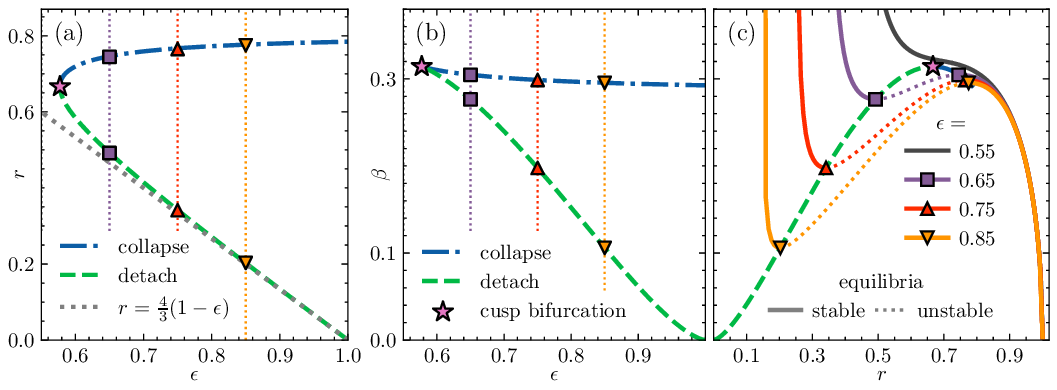}
    \caption{
    (a) Location of the bifurcations depending on the extensibility $\epsilon =\ell_{\text{max}}/\ell$.
    With increasing extensibility, the two bifurcation values of the distance $r$ between the centers of the spheres corresponding to collapse (blue dash-dotted line) and detaching (green dashed line) further separate. 
    The detaching position has the asymptotic behavior of $r = x/\ell = 4(1-\epsilon)/3$ (gray dotted line) when the extensibility approaches unity.
    At $\epsilon = 1/\sqrt{3}\approx0.6$, both curves corresponding to saddle-node bifurcations merge into a cusp bifurcation (rose star). Below this critical value of the extensibility, no bifurcations occur. 
    In (b) the magnetic field strengths $\beta$
    , at which the bifurcations occur, are shown for different extensibilities. 
    The individual bifurcation diagrams for the marked points are shown in (c), together with the curve traced out by all points of bifurcation (green dashed line). The values of the extensibilities $\epsilon = 0.65$, $0.75$, and $0.85$ for the depicted bifurcation scenarios (purple, red, and yellow, respectively) in (c) are marked in (a) and (b) for reference as vertical dotted lines of identical color.
    }
    \label{fig:ext_bif}
\end{figure*}

Dimensional analysis shows that the relevant dimensionless parameters are the extensibility $\epsilon = {\ell_{\text{max}}}/{\ell}$ of the spring and the rescaled magnetic field strength $\beta = H \sqrt{{8\pi \chi_a^2 \mu_0 a^6}/{3k \ell^5}}$. 
All lengths can be scaled by $\ell$ and energies by $k \ell^2$. The total rescaled potential for two spheres as a function of the relative separation $r = x/\ell$ then reads
\begin{equation} \label{eq:2sphere_potential}
    U_2 = -\frac{\epsilon^2}{2}\ln\left(1 - \frac{(r-1)^2}{\epsilon^2}\right) - \frac{1}{3} \frac{\beta^2}{r^3}.
\end{equation}
For small values of $\beta$, the spring potential dominates, and there is a single stable equilibrium position close to the rest length $\ell$. 
While increasing $\beta$, a second local minimum appears in a saddle-node bifurcation at small separations due to magnetic attraction. When the field strength is increased further, the first minimum disappears in another saddle-node bifurcation, and only the collapsed state remains stable. 
This behavior is illustrated in \cref{fig:pot_bif}.
The points of bifurcation depend only on the extensibility $\epsilon$ of the spring and are given as the roots of a cubic polynomial, see part A of the supporting information.
Bifurcations only appear above a critical extensibility $\epsilon = {1}/{\sqrt{3}}$. With increasing extensibility $\epsilon$, the separation between the two branches resulting from the bifurcation increases, see \cref{fig:ext_bif}(a) and (b). This concerns both separation in space $r$, see \cref{fig:ext_bif}(a), and associated field strengths $\beta$, see \cref{fig:ext_bif}(b). In \cref{fig:ext_bif}(b), the required field strength for the collapse (blue dash-dotted line) does not vary significantly with changes in extensibility. It mainly depends on the initial spring stiffness. However, as the spheres get closer, with increasing extensibility/compressibility of the springs, the magnetic field has to be lowered more until the spheres detach again (green dashed line). Individual bifurcation diagrams for different extensibilities $\epsilon$ are illustrated in \cref{fig:ext_bif}(c).

\subsection{Nonreciprocal motion of the three-sphere microswimmer}
\label{sec:non-reciprocal}
We now introduce the full three-sphere microswimmer consisting of three magnetizable beads with radius $a$, connected by FENE springs of rest lengths $\ell_{i}$, maximum extensions $\ell_{i,\text{max}}$, and stiffnesses $k_i$, where $i \in \{1,2\}$.
We assume the springs to be resistant to bending and torsion so that the spheres remain collinearly aligned. The mutual dipolar interaction together with the aligning effect of the external magnetic field support this collinear alignment, in particular during potential fabrication using methods of self-assembly \cite{holm2005structure,sanchez2015supramolecular,klapp2016collective}. If the external magnetic field is applied along the $x$-direction, the swimmer aligns itself with the field because its overall easy axis of magnetization is along its long axis. 
Hence, we only consider motion along the $x$-axis in the following.

The first spring connects the first two spheres at center locations $x_1$ and $x_2$, while the second spring connects the second to the third sphere at center locations $x_2$ and $x_3$. We denote the center-to-center distances as $x_{ij} = x_j - x_i$, where $i=1$ and $j=2$ or $i=2$ and $j=3$.
Next, we rescale all lengths by the radius $a$ and the magnetic field as $h = H/H_{\text{max}}$, where $H_{\text{max}}$ denotes the maximally attainable magnetic field strength. 
To simplify the numeric prefactors, we rescale energies by $8\pi \chi_a^2 \mu_0 a^3 H_{\text{max}}^2/3$ and the spring constants by $8\pi \chi_a^2 \mu_0 a H_{\text{max}}^2/3$.

In total, the potential of our three-sphere swimmer is then given by
\begin{eqnarray} 
    U_3 &=&{} -\frac{1}{2}{k_1\ell_{1,\text{max}}^2} \ln\left(1 - \frac{(x_{12} - \ell_1)^2}{\ell_{1,\text{max}}^2}\right) \nonumber\\&&{}- \frac{1}{2}{k_2\ell_{2,\text{max}}^2} \ln\left(1 - \frac{(x_{23} - \ell_2)^2}{\ell_{2,\text{max}}^2}\right) \nonumber \\
    &&{}- \frac{{h^2}}{3}\left( \frac{{1}}{x_{12}^3} + \frac{1}{x_{23}^3} + \frac{1}{(x_{12} + x_{23})^3}\right), \label{eq:tot_potential}
\end{eqnarray}
where all quantities are now dimensionless.
The dynamics of $x_{12}$ and $x_{23}$ are only loosely connected, because the magnetic coupling term $-h^2/[3(x_{12} + x_{23})^3]$ is comparatively small due to the larger distance between the outer spheres.  However, the presence of the additional sphere tends to reduce the equilibrium distance between each pair of spheres in a magnetized state and lowers the required field strengths for collapse and detaching. The effect is stronger when the other outer sphere is already in the collapsed state. Hence, the collapse of one spring can trigger the collapse of the other, and similarly for the expansion events. Moreover, abrupt movements of the center sphere during bifurcation events affect the spring lengths of both linked pairs of spheres. Corresponding effects are included in the following analysis of the three-sphere swimmer, while the underlying driving hysteretic effect remains the one described in Sec.~\ref{sec:hysteretic}.

We now apply a slowly oscillating magnetic field $h(t)$ that varies between a minimum magnitude $h_{\text{min}}$ and a maximum magnitude $h_{\text{max}}=1$, without changing direction. 
During one cycle of the magnetic field, both pairs of spheres will undergo hysteretic collapse and detachment at different field strengths.
The desired order of those events to achieve nonreciprocal motion and thus net propulsion of our microrobot is as follows. 
Starting from high magnetic field strengths in the collapsed state, when gradually decreasing the magnetic field, one pair of spheres separates first. Then the other pair of spheres separates. This secondary event is supported by the now weaker magnetic attraction towards the already separated outer sphere from the first separation event. When the strength of the magnetic field is increased again, the sequence of reaction is the \textit{same} for the two pairs. The pair that first separated will also be the first to collapse. Then the other pair follows.
The cycle is completed, and the swimmer is again in its initial state.

We denote the necessary rescaled magnetic fields for separation and collapse of the pair of spheres $i$ and $j$ as $h_{\text{separate},ij}$ and $h_{\text{collapse},ij}$, respectively. 
For our purpose, we require $h_{\text{separate},23} < h_{\text{separate},12} \le h_{\text{collapse},12} < h_{\text{collapse},23} \le 1$. 
It appears from here that the extensibility of the spring length $x_{23}$ should be larger. This promotes a wider gap between the field strengths of separation and collapse. 
In principle, it is not necessary for the spring length $x_{12}$ to show any hysteretic behavior at all, as long as the major deformation of this spring occurs between $h_{\text{separate},23}$ and $h_{\text{collapse},23}$.
For the magnetic field to be able to induce the full collapse, we require $h_{\text{collapse},{ij}} \lesssim 1$, which implies $k_i \ell_i^5 \lesssim 10$, see part B of the supporting information.
We therefore choose to define the stiffness parameters $c_i = k_i \ell_i^5$ for $i \in \{1,2\}$. Since these stiffness parameters are related to the strength of the magnetic field at collapse, they form a reasonable choice of parametrization adapted to the investigated situation. We found them more suitable than the pure spring constants $k_i$ when performing the optimization below.

\subsection{Hydrodynamic interactions and net propulsion}
\label{sec:hydrodynamics}

At low Reynolds numbers, the motion of the microswimmer is overdamped and governed by Stokesian hydrodynamics. Since the flow equations are linear, they lead to equations of motion for the positions of the spheres of the form
\begin{equation} \label{eq:of_motion}
    v_i = \frac{d x_i}{dt} = M_0 \sum_{j=1}^{3} M_{ij} F_j,\qquad i=1,2,3.
\end{equation}
Here, $F_j = -\partial U_3/\partial x_j$ are the forces that act on the spheres due to the potential in Eq.~(\ref{eq:tot_potential}). $M_{ij}$ are the components of the dimensionless mobility matrix $\mathbf{M}$. The latter was rescaled by the mobility of a single sphere $M_0 = {1}/{6 \pi \eta a}$, where $\eta$ is the shear viscosity of the incompressible fluid. If bending becomes significant in reality during magnetic actuation, for example, if the elastic links are substantially softer under large-scale bending than under compression, against the stabilizing collinear action of the magnetic field, or if the direction of the magnetic field is changed, \cref{eq:of_motion} must be evaluated in three dimensions. In that case, rotational degrees of freedom must be included and the three-dimensional mobility matrix including translation--rotation couplings must be evaluated \cite{babel2016dynamics}. At present, we focus on the persistently collinear case that can be reduced to one dimension.
To render this equation dimensionless and to simplify the prefactor, we rescale time by the characteristic timescale $9 \times 10^6 \eta / 4 \chi_a^2\mu_0H_{\text{max}}^2$ and velocities accordingly by $4\times 10^{-6} \chi_a^2\mu_0 H_{\text{max}}^2a/9\eta$. In these units, the mobility of a single sphere becomes $M_0 = 10^6$.

With the help of Faxén's relations, the mobility matrix can be expanded in terms of the inverse intersphere distances $1/x_{ij}$ as given in part C of the supporting information.
The expansion breaks down when the spheres are very close to each other. 
To avoid this regime, we impose a minimum separation of $4 a$ between the centers of the spheres. For this purpose, we restrict the maximum deviation from the rest length of the springs by setting $\ell_{i,\text{max}} = \ell_i - 4$, so that the FENE potential in \cref{eq:U_FENE} diverges before the distance between the spheres becomes too small. 
Furthermore, we consider the springs as frictionless. This assumption simplifies the numerical treatment. Yet, experimental results will deviate to a certain degree, depending on the exact realization of the springs. In general, through additional dampening the dynamics becomes slower such that either the driving frequency has to be reduced or the magnetic field strength and spring stiffness have to be increased to compensate for this drag. Furthermore, hydrodynamic interactions with and between the springs affect the net per-cycle displacement. To some extent, hydrodynamic friction of spring segments may be combined with that of the nearby sphere in the calculation, yet it changes between the compressed and expanded state.

Purcell's scallop theorem \cite{purcell1977life} in our case implies that the net propulsion of the swimmer after one cycle of the magnetic field only depends on the shape of the loop in configuration space $(x_{12}, x_{23})$.  
We now provide a geometric interpretation of this cycle.
For this purpose, we change to relative coordinates by transforming
\begin{equation}
    \mathbf{x} = 
    \begin{pmatrix}
        x_1 \\ x_2 \\ x_3
    \end{pmatrix}
    \rightarrow
    \underbrace{\begin{pmatrix}
        -1 & 1 & 0 \\
        0 & -1 & 1 \\
        0 & 1 & 0
    \end{pmatrix}}_{\textstyle =:\mathbf{U}}\cdot
    \begin{pmatrix}
        x_1 \\ x_2 \\ x_3
    \end{pmatrix}
    =
    \begin{pmatrix}
        x_{12} \\ x_{23} \\ x_2
    \end{pmatrix}
    = \mathbf{x}_{\text{rel}}.
\end{equation}
Introducing the vector $\vb{e} = (1,1,1)^T$ to project onto the total force, Newton's third law $\sum_{i=1}^3 F_i =: \vb{e}^T \cdot\vb{F} = 0$ can be rewritten as
\begin{equation}
    0 = \underbrace{\vb{e}^T \cdot\vb{M}^{-1} \cdot\vb{U}^{-1}}_{\textstyle :=\vb{\alpha}^T} \d \vb{x}_{\text{rel}} = \alpha_1 \d x_{12} + \alpha_2 \d x_{23} + \alpha_3 \d x_2.
\end{equation}

Now we move into the configuration space $(x_{12}, x_{23})$. 
Since the total displacement does not integrate to a function of the configuration space, we denote it as an inexact one-form $\dbar x_2$.
\begin{equation} 
    \dbar x_2 = -\frac{\alpha_1}{\alpha_3}\d x_{12} -\frac{\alpha_2}{\alpha_3}\d x_{23}.
\end{equation}
The displacement over one cycle is then given by the line integral over the configuration space loop $\gamma$,
\begin{equation}
    \Delta x_2 = \oint_\gamma \dbar x_2 = \int_{S(\gamma)} \d\dbar x_2 =: \int_{S(\gamma)} \mathcal{F},
\end{equation}
where we have used Stokes theorem to convert the line integral into a surface integral over the area $S(\gamma)$ enclosed by the loop $\gamma$.
We here defined the two-form $\mathcal{F} = \d \dbar x_2 = f \d x_{12}\wedge \d x_{23}$, together with
\begin{equation} \label{eq:curvature}
    f = \left({}-\frac{\partial}{\partial x_{12}}\left(\frac{\alpha_2}{\alpha_3}\right) + \frac{\partial}{\partial x_{23}}\left(\frac{\alpha_1}{\alpha_3}\right)\right).
\end{equation}
$f$ is the infinitesimal displacement per area in configuration space.
Introducing a configuration space metric $g$ corresponding to the dissipated energy, the scalar $f/\sqrt{\det g}$ can also be interpreted as the curvature of the configuration space \cite{avron2008geometric}. 

\begin{figure}
    \centering
    \includegraphics[width=0.998\linewidth]{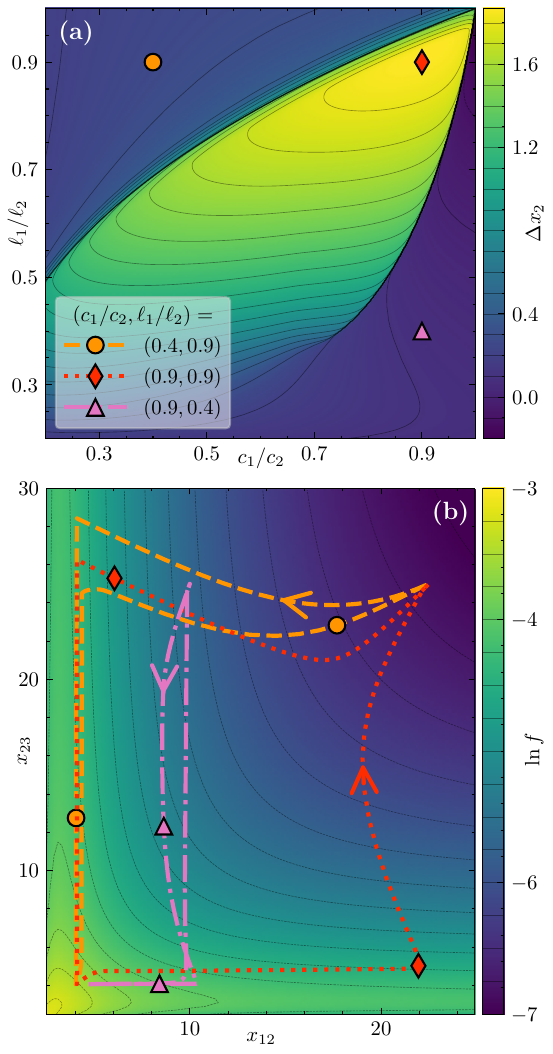}
    \caption{
    (a) Color map of the net displacement $\Delta x_2$ after one full cycle of the magnetic field for swimmers with different ratios of spring stiffness $c_1/c_2$ and rest lengths $\ell_1/\ell_2$ (phase space).
    For all swimmers $c_2 = 10$, $\ell_2 = 25$, and in the fully compressed state the center-to-center distance is still larger than $\ell_i - \ell_{i,\text{max}} = 4$. The magnetic field is ramped up from $0$ to $1$ with the very slow rate $\d h/\d t = 10^{-3}$ and then back down again to $0$. Except for the points of bifurcation, the swimmer is almost always in equilibrium.
    The three markers each label one swimmer that represents one of the three dynamical states in this phase space. Their trajectories in configuration space are shown in (b).
    (b) Color map of the natural logarithm of the infinitesimal displacement $f$ per area in configuration space $(x_{12}, x_{23})$ calculated according to \cref{eq:curvature} from the sixth-order expansion of the mobility matrix \cref{eq:mobility_matrix}. 
    Three loops in configuration space traced out during one cycle of the magnetic field by the three-sphere swimmers with the different parameters as marked in (a).
    }
    \label{fig:conf_space}
\end{figure}

As can be seen in the phase space \cref{fig:conf_space}~(a), there are three distinct dynamical states. 
In the state that achieves the highest net displacement (red diamond), the collapses occur in the correct sequence described above, leading to a large loop in configuration space, see \cref{fig:conf_space}~(b). To achieve this extended loop, it is essential that the springs notably differ both in rest length and in stiffness.
If, from there, in \cref{fig:conf_space}~(a) the stiffness ratio $c_1/c_2$ is decreased (orange circle), the detachment of the first pair of spheres occurs after the separation of the second pair. The enclosed area in configuration space is much smaller, and the net displacement decreases significantly.
If, instead, the stiffness ratio $c_1/c_2$ is kept constant and the rest length ratio $\ell_1/\ell_2$ is decreased (pink triangle), the collapse of the second pair occurs earlier than that of the first pair. The consequence is again a much smaller loop in configuration space, together with a significantly lower, sometimes even negative, net displacement.
Between the circle and the diamond, there is a relatively sharp transition between the different dynamic states. The transition between these states apparently vanishes at a critical point when $c_1/c_2 \approx 0.75$ and $\ell_1/\ell_2 \approx 0.4$. 

For smaller length ratios, the length of the first spring $l_1 < 10$ is too short, and the extensibility is below $\epsilon_1 = \ell_{1,\text{max}}/\ell_1 < 0.6$.
As shown in Sec.~\ref{sec:hysteretic} for two spheres, no bifurcations, and thus no hysteretic behavior occurs for such low extensibilities. Due to its absence, the transition is thus more continuous.

\subsection{Evolutionary optimization of swimmer design and driving field}
\label{sec:evolution}

The design of the swimmer can be optimized to achieve maximum swimming speed under practical constraints, see Methods. To ensure real-world applicability, we now choose dimensional parameters. 
The spheres of radius $a = \SI{1}{\micro\meter}$ are made of a superparamagnetic material with high magnetic susceptibility $\chi \gg 1$, suspended at room temperature in water with dynamic viscosity $\eta = \SI{7e-4}{\pascal\second}$ and density $\rho = \SI{1000}{kg/m^3}$. 
Generally, higher external magnetic fields lead to stronger induced magnetic forces and thus larger speeds of the spheres, relating to the overall speed of the swimmer. Simultaneously, experimental realizability and required safety of in-vivo applications impose constraints on the field strength.
Hence, for the following numeric study, the magnetic field is restricted in magnitude by $H_{\text{max}} = \SI{1e3}{Oe} = \SI{100}{\milli\tesla}/\mu_0 $ and can be varied at a maximum rate of $|\d H/ \d t | \le \SI{2.7e4}{Oe\per\second}$ following the ICNIRP guidelines \cite{icnirp2014guidelines} to remain safe for in vivo applications. Thus, using the dimensionless $h$ and $t$ as introduced in Secs.~\ref{sec:non-reciprocal} and \ref{sec:hydrodynamics}, $|\d h/\d t| \le 0.59$.
A simple driving magnetic field takes the form of a cosine wave that oscillates between $h_{\text{min}}$ and $h_{\text{max}}$ with angular frequency $\omega$. In this case, the constraints on the field magnitudes translate to $|h_{\text{max}}| 
\le 1
$, $|h_{\text{min}}| \le 1$, and $\omega |h_{\text{max}} - h_{\text{min}}| / 2 \le 0.59$.
For the swimmer, we restrict the length $\ell_i - \ell_{i,\text{max}} \geq 4$ to guarantee that the surfaces of the spheres always remain at least one diameter apart. Thus we avoid lubrication effects and reduce mutual magnetization. Moreover, we choose the spring stiffness $c_i = k_i \ell_i^5$ in the range $1 \le c_i \le 20$, and we set the rest lengths $5 \le \ell_i \le 50$.

\begin{figure}
    \centering
    \includegraphics[width=\linewidth, trim=10 0 10 0, clip]{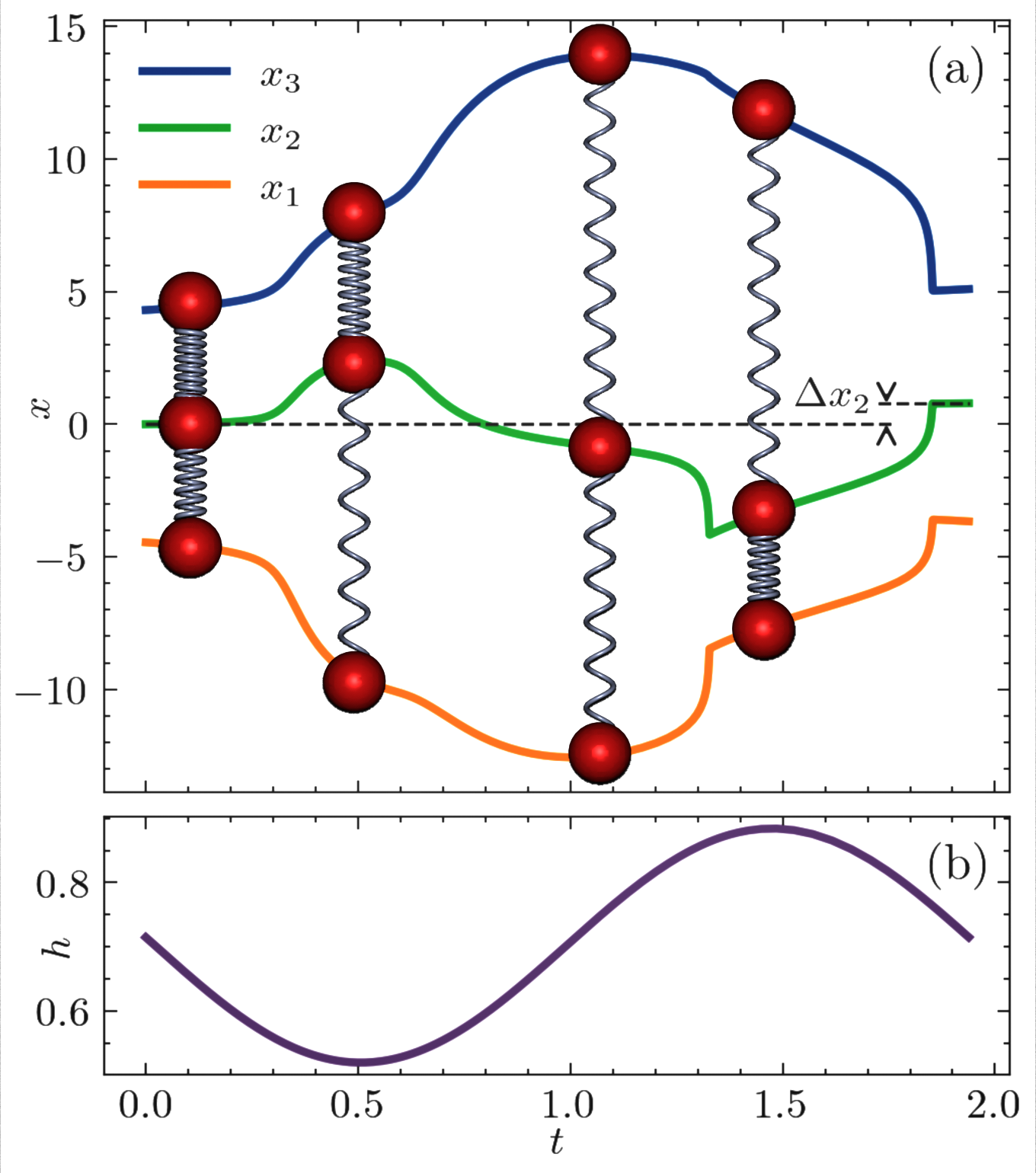}
    \caption{
    (a) One cycle of the trajectories of a three-sphere microswimmer with optimized parameters $c_1=6.63$, $c_2=8.22$, $\ell_1=13.7$, $\ell_2=17.0$ and $\ell_{i,\text{max}} = \ell_i - 4$. 
    The net displacement accrued over one cycle of the magnetic field is $\Delta x_2 = 0.791$ in dimensionless units. This gives an average speed of $v = 0.408$ over the full cycle or $v = \SI{18.6}{\micro\meter\per\second}$ in physical units when using the scales described in Sec.~\ref{sec:evolution}. In each of the four stages of the cycle, a snapshot of the swimmer consisting of three spheres (red) and two springs (silver-blue) is shown, where the spheres are depicted to scale. 
    (b) The driving magnetic field $h(t)$ oscillates between $h_{\text{min}} = 0.52$ and $h_{\text{max}} = 0.884$ with angular frequency $\omega = 3.24$. 
    }
    \label{fig:cos_traj}
\end{figure}

Even though gradients are available through automatic differentiation, gradient-based optimization approaches were not successful. This is because the optimization landscape is highly nonconvex and contains sharp cliffs at the transitions between different dynamics states, which can be seen in \cref{fig:conf_space}~(a). 
We thus optimize this $7$-dimensional problem over $(\ell_1, \ell_2, c_1, c_2, h_{\text{min}}, h_{\text{max}}, \omega)$ using the covariance matrix adaptation evolutionary strategy (CMA ES) \cite{hansen2006cma} with a population size of $32$ for $150$ generations, until the solution stops improving. The final solution is shown in \cref{fig:cos_traj}. 
In physical units, this swimmer of total outer length on the order of $\SI{33}{\micro\meter}$ swims at an average speed of $\SI{18.6}{\micro\meter\per\second}$ when driven by a magnetic field oscillating between $\SI{520}{Oe}$ and $\SI{884}{Oe}$ at a frequency of $\SI{23.5}{\hertz}$. The spring stiffnesses resulting from the optimization process are $k_1 = \SI{1.2e-4}{N/m}$ and $k_2 = \SI{4.1e-5}{N/m}$. 

Due to the constraints imposed on the slew rate of the magnetic field $\d h/\d t$, the optimized driving field does not drop to zero but stays in the interval where hysteretic effects are important. This, in turn, allows for the frequency to be increased, which improves the speed. Besides, the optimized driving field never reaches the maximum allowed strength ($1$ in rescaled units). Thus, using stronger electromagnets while maintaining the restriction on the slew rate would not further enhance the overall swimming speed.

Since the driving frequency is much faster than in the quasistatic case described in Sec.~\ref{sec:hydrodynamics}, the detaching and collapse do not happen instantaneously, but with significant delay. 
The swimmer reaches its fully collapsed state only after the magnetic field has reached maximum amplitude and is already declining again. 
During this cycle, the maximum speed of a single sphere $v_{\text{max}} = \SI{1.97e-2}{m/s}$ is reached during the collapse of the first spring at $t=1.72$. Thus, the Reynolds number is
\begin{align}
    \text{Re} = \frac{2\rho v_{\text{max}} a}\eta =\num{5.7e-2} \ll 1,
\end{align}
so that the assumption of Stokes flow still remains approximately valid. 

\begin{figure}
    \centering
    \includegraphics[width=\linewidth]{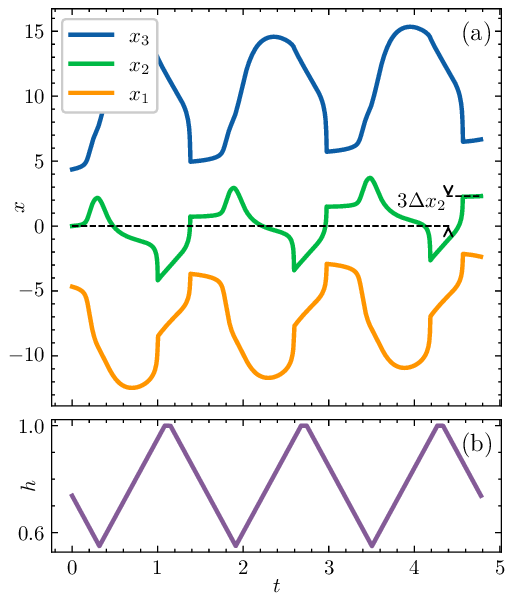}
    \caption{
    (a) Three cycles of the trajectories of a three-sphere microswimmer with parameters $c_1=8.1$, $c_2=9.8$, $\ell_1=13.2$, $\ell_2=17.0$ and $\ell_{i,\text{max}} = \ell_i - 4$. 
    The net displacement accrued over one cycle of the magnetic field is $\Delta x_2 = 0.769$. This gives an average speed of $v = 0.483$ over the full cycle or $v = \SI{22.0}{\micro\meter\per\second}$ in physical units when using the scales described in Sec.~\ref{sec:evolution}.
    (b) Associated driving magnetic field $h(t)$ of triangular shape with constant slew rate $\d h/ \d t = \pm 0.59$ between $h_{\text{min}} = 0.55$ and $h_{\text{max}} = 1.02$, which has then been clipped to the interval $[0,1]$. The period is thus $1.59$. 
    One unit of time on the abscissa corresponds to $\SI{2.2e-2}{s}$.
    }
    \label{fig:bangbang_traj}
\end{figure}

If we allow for more general shapes of the forcing magnetic field $h(t)$, the velocity can be further increased. The term $h^2(t)$ enters affinely into the total force, resulting from the potential in Eq.~(\ref{eq:tot_potential}). Consequently, by Pontryagin's maximum principle \cite{pontryagin2018mathematical} from optimal control theory, the magnetic field which leads to the extremal velocity must lie on the boundary of the manifold, that is, be a \textit{bang-bang control}. Thus,  at each point in time $t$, either $h(t) \in \{0,1\}$ or $\d h/\d t = \pm 0.59$ locates extrema, according to the bounds of optimization introduced above. 
Therefore, a reasonable class of driving fields is given by triangle wave functions clipped to the interval $[0,1]$. Performing the same evolutionary optimization strategy within this class gives rise to a swimmer whose cycles are shown in \cref{fig:bangbang_traj}. The amplitude of the magnetic field in \cref{fig:bangbang_traj}~(b) remains similar to the one of the cosine-shaped field in \cref{fig:cos_traj}~(b). Yet, the frequency can be increased whilst keeping similar displacement per period and thus the swimmer is around $18\%$ faster.

\begin{figure}
    \centering
    \includegraphics[width=\linewidth]{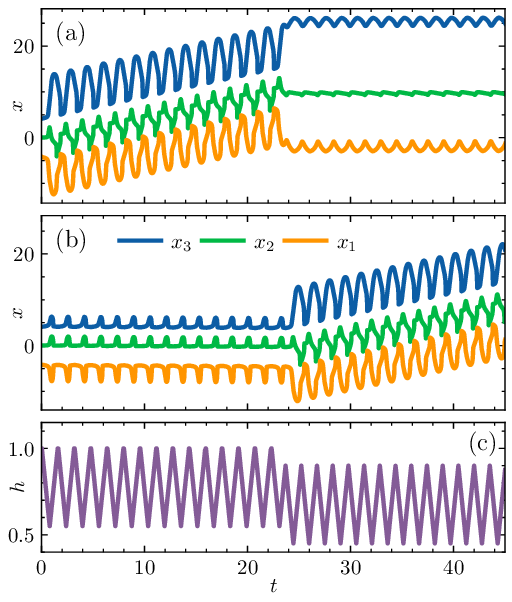}
    \caption{
    (a) The trajectories of a three-sphere microswimmer with parameters $c_1=8.1$, $c_2=9.8$, $\ell_1=13.2$, $\ell_2=17.0$ and $\ell_{i,\text{max}} = \ell_i - 4$, identically to \cref{fig:bangbang_traj}. At first, when the driving magnetic field has the right magnitude, the swimmer completes full cycles leading to net displacement. When the driving field is decreased in magnitude, the swimmer does not reach the fully collapsed state and by lack of pronounced nonreciprocity it oscillates in place.
    (b) Setting the parameters of a second three-sphere microswimmer differently, here to $c_1=6.0,$ $c_2=6.9$, $\ell_1=13.0$, $\ell_2=17.0$, and $\ell_{i,\text{max}} = \ell_i - 4$, can facilitate targeted stimulation of one of the swimmers by adjusting the field magnitudes. First, when the driving magnetic field has a larger magnitude, since the springs are weaker, the swimmer never fully expands, and the cycles remain incomplete. The swimmer oscillates basically in place. When the driving field is decreased in magnitude, the swimmer completes full cycles and therefore achieves substantial net displacement.
    (c) Throughout, the driving magnetic field $h(t)$ of triangular shape is of fixed slew rate $\d h/ \d t = \pm 0.59$. Yet, the magnitude switches at one point. For the first $15$ periods, the field oscillates between $h_{\text{min}}^{(1)} = 0.55$ and $h_{\text{max}}^{(1)} = 1.02$, clipped to the interval $[0,1]$. Then, for another 15 periods, it oscillates between $h_{\text{min}}^{(2)} = 0.45$ and $h_{\text{max}}^{(2)} = 0.9$. 
    One unit of time on the abscissa corresponds to $\SI{2.2e-2}{s}$.
    }
    \label{fig:indiv_control}
\end{figure}

If a different driving field is used, which oscillates too rapidly or does not reach a high enough maximum or low enough minimum during one period, the swimmer does not complete an entire cycle of both collapses and detachments, so that it cannot achieve significant net displacement. This dependence can be used to independently control multiple swimmers individually, as shown in \cref{fig:indiv_control}. 
There, appropriate adjustment of the maximum and minimum field magnitudes allows to either address one or the other of the two swimmers in a way to induce substantial net motion. In contrast to that, the second swimmer remains basically at rest during that time.

\section{Conclusion and Outlook}
\label{sec:conclusion}
We have shown that a simple microswimmer consisting of three magnetizable spheres connected by two springs can achieve net propulsion when driven by an oscillating magnetic field, even at low Reynolds numbers. 
The key ingredient is the hysteretic collapse and detaching of the spheres due to bifurcations of the equilibrium positions induced by the competition of magnetic attraction and elastic counteraction by the springs.
Due to the simple design of the swimmer, a straightforward description of its dynamics is possible by expanding hydrodynamic interactions in terms of the inverse intersphere distances. 
Using an evolutionary optimization strategy, we have improved the swimmer design and driving magnetic field to achieve high swimming speeds under practical constraints. 
We demonstrate that a microrobot made out of superparamagnetic spheres of radius $\SI{1}{\micro\meter}$, connected by two springs with stiffnesses on the order of $\SI{1e-4}{N/m}$ with rest lengths of around $\SI{13}{\micro\meter}$ and $\SI{17}{\micro\meter}$ could reach swimming speeds of around $\SI{20}{\micro\meter\per\second}$ when actuated by an oscillating magnetic field of magnitude $\SI{1000}{Oe}$ and frequencies of approximately $\SI{25}{\hertz}$.

Artificial bacteria flagella (ABF) \cite{nelson2009abf}, a promising alternative design for a magnetically actuated microswimmer, can achieve similar speeds ($\SI{18}{\micro\meter\per\second}$) at a comparable scale ($\SI{16}{\micro\meter}$ length and $\SI{5}{\micro\meter}$ radius helix) \cite{ding2016microfluidic}. 
However, they require a different kind of magnetic actuation, that is, a rotating magnetic field, in Ref.~\cite{nelson2009abf} with an amplitude of $\SI{9}{mT}$. Newer designs using hard magnetic materials can reach speeds on the order of millimeters per second \cite{wu2025femtosecond} due to higher frequencies of rotating fields. Differently shaped ABFs can be addressed independently by tuning the actuation frequency \cite{amoudruz2022independent}, while we in our case switch the field magnitudes to selectively actuate our different swimmers.
Our approach uses an order of magnitude stronger magnetic fields but of a simpler geometry, oscillating in magnitude with constant direction. This could make actuation and control more accessible. However, the fabrication of such a three-sphere swimmer may prove to be more challenging than that of artificial bacterial flagella because of the moving parts and the need for appropriate elastic springs. 

The proposed design could be realized using superparamagnetic iron oxide \cite{dulinska2019superparamagnetic}, DNA springs \cite{iwaki2016programmable, goubault2003flexible}, or 3D printed elastomers \cite{eom2025fast} as springs. To produce the strong and rapidly oscillating uniform driving field, a set of liquid-cooled high-performance Maxwell coils could be employed. Alternatively, the magnitude of the driving field can be scaled down together with the spring stiffnesses at the cost of a slower driving frequency and in turn lower swimming speeds. Our employed optimization strategy and some expected adjustability of the magnitudes and frequencies of the external magnetic field can help to mitigate deviations during experimental realization from the calculated theoretical design parameters.
Evolutions of this design, which go beyond the simple minimalist model presented here, could include a microrobot that is entirely printed from an elastomer with embedded magnetic particles that deform the robot body under the influence of an external magnetic field. 
More complex motion patterns, such as beating cilia or undulating sperm tails, could be achievable. We hope that this theoretical and numerical study can inspire future experimental realizations and investigations of such microswimmers.

\newpage

\section{Methods}

The equations of motion, \cref{eq:of_motion}, can be numerically integrated using the implicit Kvaerno5 solver \cite{kvaerno2004singly} with adaptive time stepping \cite{hairer2008solving-i} and absolute tolerance $\num{1e-6}$. An implicit solver is crucial because of the stiffness of the problem.
Benefiting from just-in-time compilation and hardware acceleration, the program was implemented using the JAX ecosystem \cite{jax2018github} in the programming language Python. Notably, we used the libraries Diffrax \cite{kidger2021on}, Equinox \cite{kidger2021equinox}, and Evosax \cite{evosax2022github}. We used the numerical integration of \cref{eq:of_motion} to provide the net displacement per cycle in \cref{fig:conf_space}(a) and the loops in the configuration space presented in \cref{fig:conf_space}(b).

\section*{Associated Content}
\subsection{Data Availability Statement}
All the code and data to reproduce the figures are available at \url{https://github.com/TheoLequy/magspheres}.

\subsection{Supporting Information}
The supporting information in part A details the analytical derivation of the positions of bifurcation shown in \cref{fig:ext_bif}.
Second, in part B, it estimates the magnitude of the spring stiffnesses for the three-sphere swimmer to exploit the available magnitude of the external magnetic field. And third, in part C, it lists the sixth-order expressions of the mobility matrix introduced in \cref{eq:of_motion}.

\begin{acknowledgments}
T.L.\ thanks the Studienstiftung des Deutschen Volkes (German Academic Scholarship Foundation), Deutscher Akademischer Austauschdienst (German Academic Exchange Service), and the Excellence Scholarship \& Opportunity Programme (ESOP) from ETH Zürich for financial and academic support.
A.M.M.\ acknowledges support by the Deutsche Forschungsgemeinschaft (German Research Foundation, DFG) through project no.\ 541972050, DFG reference no.~ME 3571/12-1. 
\end{acknowledgments}

\newpage

\section*{Supporting Information}

\renewcommand{\theequation}{S\arabic{equation}}
\setcounter{equation}{0}
\setcounter{subsection}{0}

\subsection{Analytical expressions for bifurcation locations}
\label{sec:analytical}

The equilibrium positions of the two-sphere setup are found from 
$\partial U_2/\partial r=0$, see \cref{eq:2sphere_potential}.
From there, we obtain a quintic polynomial equation in $r$,
\begin{equation} \label{eq:beta_bif}
    0 = r^5 - r^4 + \frac{\beta^2}{\epsilon^2} \left(\epsilon^2 - (r-1)^2\right)\,.
\end{equation}
However, among the resulting roots, at most three are physical with $r \in (1-\epsilon, 1+\epsilon)$.

Saddle-node bifurcations occur at the local extrema of $\beta$. Setting $\d \beta/\d r = 0$, 
we obtain a cubic polynomial equation,
\begin{equation}
    0 = 3r^3 - 10 r^2 + (11 - 5 \epsilon^2) r + 4(\epsilon^2 - 1)\,,
\end{equation}
with discriminant
\begin{equation}
    \Delta = 4 \epsilon^2 (3\epsilon^2 - 1) (125 \epsilon^2 + 1)\,.
\end{equation}
Only for $\epsilon \in ({1}/{\sqrt{3}}, 1]$ the discriminant is positive and two physical bifurcations occur.
At $\epsilon = 1/\sqrt{3}$, both bifurcations merge at $r=2/3$ and $\beta=2\sqrt{2}/9$.

\subsection{Magnitude of the spring stiffnesses of the three-sphere swimmer}

As can be inferred from Fig.~3(b) of the main article, the strength of the magnetic field for collapse (blue dash-dotted line) does not significantly vary with extensibility $\epsilon$, so that we can approximate
\begin{equation}
    \label{eq:betacollapse}
    \beta_{\text{collapse}} \sim 0.3\,.
\end{equation}
We use this result from the two-sphere dynamics, which describes the basic underlying hysteretic effect, for our following estimate for the three-sphere case. That is, we only consider the two spheres $(i,j) = (1,2)$ or $(2,3)$ and ignore the weaker influence of the third sphere, which is significantly further away from the outer sphere than the central one. To transform from the rescaled dimensionless strength of the magnetic field $\beta$ in Sec.~II~\ref{sec:analytical} to the magnetic field parameter $h$, we use the definition of $\beta$ introduced before \cref{eq:2sphere_potential} in the main text. For each of the two mentioned pairs of spheres $(i,j)$, this definition for the rescaled magnetic field at collapse can be rewritten as
\begin{align}
    \beta_{\text{collapse},ij} &= H_{\text{collapse}, ij}\sqrt{{8\pi} \chi_a^2 \mu_0 {a^6}/{3k_i\ell_i^5}} \nonumber \\ 
    &= \frac{H_{\text{collapse}, ij}}{H_{\text{max}}} \sqrt{\frac{8\pi\chi_a^2 \mu_0 a H^2_{\text{max}}}{3 k_i} \frac{a^5}{\ell_i^5}}\,.
    \label{eq:rescaling}
\end{align}
In this expression, we recognize our rescaled quantities as introduced before Eq.~(\ref{eq:tot_potential}) in the main text. Spring constants $k_i$ are measured in units of $8\pi\chi_a^2 \mu_0 a H^2_{\text{max}}/3$, rest lengths $\ell_i$ in units of $a$, and the magnetic field in units of $H_{\text{max}}$, where we defined $h = H/H_{\text{max}}$.

Together, we infer from \cref{eq:rescaling} in rescaled units the relation
\begin{equation}
    h_{\text{collapse}, ij} = \beta_{\text{collapse}, ij} \sqrt{k_i\ell_i^5}\,.
\end{equation}
In combination with \cref{eq:betacollapse} and in terms of the stiffness parameters $c_i = k_i\ell_i^5$ introduced in Sec.~II~B, we therefore obtain
\begin{equation}
    h_{\text{collapse}, ij} \sim 0.3 \sqrt{c_i}\,.
\end{equation}
Thus, if we require $h_{\text{collapse},ij} \sim 1$, we find $c_i \sim 10$.

\subsection{Mobility matrix}
\label{sec:mob_mat}

In Ref.~\onlinecite{MobMatrix2019}, expressions for the displaceability matrix for spherical particles in a linearly elastic medium were derived up to the sixth order in the inverse particle separation distance. 
These results become equivalent to those for Stokes flow when we take the limit of $\nu\rightarrow0.5$ for the Poisson ratio. 
Utilizing the one-dimensionality of our setup, every entry in the mobility matrix becomes a scalar. Then, the diagonal and off-diagonal entries read
\begin{subequations} \label{eq:mobility_matrix}
\begin{align} 
    M_{ii} &= 1 + \sum_{k\neq i}\left[ -3.75 x_{ik}^{-4} + 5.5 x_{ik}^{-6}\right], \\
    M_{ij, i\neq j} &= 1.5 |x_{ij}|^{-3} + \biggl[-3.75 \sigma_{ij} x_{ik}^{-2}x_{jk}^{-2} \nonumber\\ 
    & + 6\sigma_{ij}\left(x_{ik}^{-2}x_{jk}^{-4} +  x_{ik}^{-4}x_{jk}^{-2}\right) - 6.5 |x_{ik}^{-3}x_{jk}^{-3}| \biggr]_{k\neq i,j},
\end{align}
\end{subequations}
respectively. 
The sign of some terms is flipped if the $k$th sphere is located in between spheres $i$ and $j$, according to  
\begin{equation}
\sigma_{ij} = \begin{cases}
    -1, & \text{if } \{i,j\} = \{1,3\},\\ 
    +1, & \text{otherwise}\,.
\end{cases}
\end{equation}
Since, in the main text, we limit the maximum approach of two spheres to a surface-to-surface distance of one diameter of the spheres, truncation beyond the sixth order leads to only little quantitative deviations.

\bibliography{references}

@article{mag1994manip,
    author = {Gillies,G. T.  and Ritter,R. C.  and Broaddus,W. C.  and Grady,M. S.  and Howard,M. A.  and McNeil,R. G. },
    title = {Magnetic manipulation instrumentation for medical physics research},
    journal = {Review of Scientific Instruments},
    volume = {65},
    number = {3},
    pages = {533-562},
    year = {1994},
    }

@article{shen2023magnetically,
  title={Magnetically driven microrobots: Recent progress and future development},
  author={Shen, Honglin and Cai, Shuxiang and Wang, Zhen and Ge, Zhixing and Yang, Wenguang},
  journal={Materials \& Design},
  volume={227},
  pages={111735},
  year={2023},
  publisher={Elsevier}
}

@article{elgeti2015physics,
  title={Physics of microswimmers—single particle motion and collective behavior: a review},
  author={Elgeti, Jens and Winkler, Roland G and Gompper, Gerhard},
  journal={Reports on Progress in Physics},
  volume={78},
  number={5},
  pages={056601},
  year={2015},
  publisher={IOP Publishing}
}

@article{avron2008geometric,
  title={A geometric theory of swimming: {P}urcell's swimmer and its symmetrized cousin},
  author={Avron, Joseph E and Raz, Oren},
  journal={New Journal of Physics},
  volume={10},
  number={6},
  pages={063016},
  year={2008},
  publisher={IOP Publishing}
}

@article{nelson2009abf,
    author = {Zhang, Li  and Abbott, Jake J.  and Dong, Lixin  and Kratochvil, Bradley E.  and Bell, Dominik  and Nelson, Bradley J. },
    title = {Artificial bacterial flagella: Fabrication and magnetic control},
    journal = {Applied Physics Letters},
    volume = {94},
    number = {6},
    pages = {064107},
    year = {2009}
    }

@Article{puljiz2018reversible,
    author ={Puljiz, Mate and Huang, Shilin and Kalina, Karl A. and Nowak, Johannes and Odenbach, Stefan and Kästner, Markus and Auernhammer, Günter K. and Menzel, Andreas M.},
    title  ={Reversible magnetomechanical collapse: virtual touching and detachment of rigid inclusions in a soft elastic matrix},
    journal  ={Soft Matter},
    year  ={2018},
    volume  ={14},
    issue  ={33},
    pages  ={6809-6821},
    publisher  ={The Royal Society of Chemistry},
    doi  ={10.1039/C8SM01051J},
    url  ={http://dx.doi.org/10.1039/C8SM01051J}
}

@article{najafi2004simple,
  title = {Simple swimmer at low {R}eynolds number: three linked spheres},
  author = {Najafi, Ali and Golestanian, Ramin},
  journal = {Physical Review E},
  volume = {69},
  issue = {6},
  pages = {062901},
  numpages = {4},
  year = {2004},
  month = {Jun},
  publisher = {American Physical Society},
  doi = {10.1103/PhysRevE.69.062901},
  url = {https://link.aps.org/doi/10.1103/PhysRevE.69.062901}
}

@article{babel2016dynamics,
    doi = {10.1209/0295-5075/113/58003},
    url = {https://dx.doi.org/10.1209/0295-5075/113/58003},
    year = {2016},
    month = {mar},
    publisher = {EDP Sciences, IOP Publishing and Società Italiana di Fisica},
    volume = {113},
    number = {5},
    pages = {58003},
    author = {S. Babel and H. Löwen and A. M. Menzel},
    title = {Dynamics of a linear magnetic “microswimmer molecule”},
    journal = {Europhysics Letters}
}

@article{jang2019targeted,
  title={Targeted drug delivery technology using untethered microrobots: a review},
  author={Jang, Deasung and Jeong, Jinwon and Song, Hyeonseok and Chung, Sang Kug},
  journal={Journal of Micromechanics and Microengineering},
  volume={29},
  number={5},
  pages={053002},
  year={2019},
  publisher={IOP Publishing}
}

@article{nelson2023delivering,
  title={Delivering drugs with microrobots},
  author={Nelson, Bradley J and Pan{\'e}, Salvador},
  journal={Science},
  volume={382},
  number={6675},
  pages={1120--1122},
  year={2023},
  publisher={American Association for the Advancement of Science}
}

@article{palagi2019light,
  title={Light-controlled micromotors and soft microrobots},
  author={Palagi, Stefano and Singh, Dhruv P and Fischer, Peer},
  journal={Advanced Optical Materials},
  volume={7},
  number={16},
  pages={1900370},
  year={2019},
  publisher={Wiley Online Library}
}

@article{schrage2023ultrasound,
  title={Ultrasound microrobots with reinforcement learning},
  author={Schrage, Matthijs and Medany, Mahmoud and Ahmed, Daniel},
  journal={Advanced Materials Technologies},
  volume={8},
  number={10},
  pages={2201702},
  year={2023},
  publisher={Wiley Online Library}
}

@book{pontryagin2018mathematical,
  title={Mathematical {T}heory of {O}ptimal {P}rocesses},
  author={Pontryagin, Lev Semenovich},
  year={2018},
  publisher={Routledge},
  address={London}
}

@article{MobMatrix2019,
  title = {Displacement field around a rigid sphere in a compressible elastic environment, corresponding higher-order {F}ax\'en relations, as well as higher-order displaceability and rotateability matrices},
  author = {Puljiz, Mate and Menzel, Andreas M.},
  journal = {Physical Review E},
  volume = {99},
  issue = {5},
  pages = {053002},
  numpages = {18},
  year = {2019},
  month = {May},
  publisher = {American Physical Society},
  doi = {10.1103/PhysRevE.99.053002},
  url = {https://link.aps.org/doi/10.1103/PhysRevE.99.053002}
}

@article{daddi2018state,
  title={State diagram of a three-sphere microswimmer in a channel},
  author={Daddi-Moussa-Ider, Abdallah and Lisicki, Maciej and Mathijssen, Arnold JTM and Hoell, Christian and Goh, Segun and B{\l}awzdziewicz, Jerzy and Menzel, Andreas M and L{\"o}wen, Hartmut},
  journal={Journal of Physics: Condensed Matter},
  volume={30},
  number={25},
  pages={254004},
  year={2018},
  publisher={IOP Publishing}
}

@article{daddi2018swimming,
  title={Swimming trajectories of a three-sphere microswimmer near a wall},
  author={Daddi-Moussa-Ider, Abdallah and Lisicki, Maciej and Hoell, Christian and L{\"o}wen, Hartmut},
  journal={Journal of Chemical Physics},
  volume={148},
  number={13},
  year={2018},
  publisher={AIP Publishing}
}

@article{annunziata2013ferrogel,
    author = {Annunziata, Mario Alberto and Menzel, Andreas M. and Löwen, Hartmut},
    title = "{Hardening transition in a one-dimensional model for ferrogels}",
    journal = {Journal of Chemical Physics},
    volume = {138},
    number = {20},
    pages = {204906},
    year = {2013},
    month = {05},
    issn = {0021-9606},
    doi = {10.1063/1.4807003},
    url = {https://doi.org/10.1063/1.4807003},
}

@article{warner1972kinetic,
  title={Kinetic theory and rheology of dilute suspensions of finitely extendible dumbbells},
  author={Warner, Harold R Jr.},
  journal={Industrial \& Engineering Chemistry Fundamentals},
  volume={11},
  number={3},
  pages={379--387},
  year={1972},
  publisher={ACS Publications}
}

@article{goh2018dynamics,
  title={Dynamics in a one-dimensional ferrogel model: relaxation, pairing, shock-wave propagation},
  author={Goh, Segun and Menzel, Andreas M and L{\"o}wen, Hartmut},
  journal={Physical Chemistry Chemical Physics},
  volume={20},
  number={22},
  pages={15037--15051},
  year={2018},
  publisher={Royal Society of Chemistry}
}

@article{leoni2009basic,
  title={A basic swimmer at low {R}eynolds number},
  author={Leoni, Marco and Kotar, Jurij and Bassetti, Bruno and Cicuta, Pietro and Lagomarsino, Marco Cosentino},
  journal={Soft Matter},
  volume={5},
  number={2},
  pages={472--476},
  year={2009},
  publisher={Royal Society of Chemistry}
}

@article{grosjean2016realization,
  title={Realization of the {N}ajafi-{G}olestanian microswimmer},
  author={Grosjean, Galien and Hubert, Maxime and Lagubeau, Guillaume and Vandewalle, Nicolas},
  journal={Physical Review E},
  volume={94},
  number={2},
  pages={021101},
  year={2016},
  publisher={APS}
}

@article{purcell1977life,
  title={Life at low {R}eynolds numbers},
  author={Purcell, E. M.},
  journal={American Journal of Physics},
  volume={45},
  number={1},
  pages={3--11},
  year={1977},
  publisher={}
}

@article{avron2005pushmepullyou,
  title={Pushmepullyou: an efficient micro-swimmer},
  author={Avron, Joseph E and Kenneth, Oded and Oaknin, David H},
  journal={New Journal of Physics},
  volume={7},
  number={1},
  pages={234},
  year={2005},
  publisher={IOP Publishing}
}

@article{biller2014modeling,
  title={Modeling of particle interactions in magnetorheological elastomers},
  author={Biller, AM and Stolbov, OV and Raikher, Yu L},
  journal={Journal of Applied Physics},
  volume={116},
  number={11},
  year={2014},
  publisher={AIP Publishing}
}

@article{biller2015mesoscopic,
  title={Mesoscopic magnetomechanical hysteresis in a magnetorheological elastomer},
  author={Biller, AM and Stolbov, OV and Raikher, Yu L},
  journal={Physical Review E},
  volume={92},
  number={2},
  pages={023202},
  year={2015},
  publisher={APS}
}

@article{icnirp2014guidelines,
  title={Guidelines for limiting exposure to electric fields induced by movement of the human body in a static magnetic field and by time-varying magnetic fields below 1 {Hz}},
  author = {{International Commission on Non-Ionizing Radiation Protection}},
  journal={Health Physics},
  volume={106},
  number={3},
  pages={418--425},
  year={2014}
}

@phdthesis{kidger2021on,
    title={{O}n {N}eural {D}ifferential {E}quations},
    author={Patrick Kidger},
    year={2021},
    school={University of Oxford},
}

@article{kidger2021equinox,
    author={Patrick Kidger and Cristian Garcia},
    title={{E}quinox: neural networks in {JAX} via callable {P}y{T}rees and
           filtered transformations},
    year={2021},
    journal={Differentiable Programming workshop at Neural Information Processing
             Systems},
    note={{A}vailable online at: \url{https://docs.kidger.site/equinox/} (Retrieved: 2025-12-05)}
}

@misc{jax2018github,
  author = {James Bradbury and Roy Frostig and Peter Hawkins and Matthew James Johnson
            and Chris Leary and Dougal Maclaurin and George Necula and Adam Paszke and
            Jake Vander{P}las and Skye Wanderman-{M}ilne and Qiao Zhang},
  title = {{JAX}: composable transformations of {P}ython+{N}um{P}y programs},
  note = {{A}vailable online at: \url{http://github.com/google/jax}, version 0.8.1 (Retrieved: 2025-12-05)},
  url = {http://github.com/google/jax},
  version = {0.8.1},
  year = {2018},
}

@article{kvaerno2004singly,
  title={Singly diagonally implicit {R}unge--{K}utta methods with an explicit first
         stage},
  author={Kv{\ae}rn{\o}, Anne},
  journal={BIT Numerical Mathematics},
  volume={44},
  number={3},
  pages={489--502},
  year={2004},
  publisher={Springer}
}

@book{hairer2008solving-i,
  address={Berlin},
  author={Hairer, E. and N{\o}rsett, S.P. and Wanner, G.},
  edition={Second Revised},
  publisher={Springer},
  title={{S}olving {O}rdinary {D}ifferential {E}quations {I} {N}onstiff
         {P}roblems},
  year={2008}
}

@article{evosax2022github,
    author  = {Robert Tjarko Lange},
    title   = {evosax: {JAX}-based Evolution Strategies},
    journal = {arXiv preprint arXiv:2212.04180},
    year    = {2022},
}

@inbook{hansen2006cma,
    author={Hansen, Nikolaus},
    editor={Lozano, Jose A. and Larra{\~{n}}aga, Pedro and Inza, I{\~{n}}aki and Bengoetxea, Endika},
    chapter={The {CMA} Evolution Strategy: A Comparing Review},
    title={Towards a {N}ew {E}volutionary {C}omputation: {A}dvances in the {E}stimation of {D}istribution {A}lgorithms},
    year={2006},
    publisher={Springer},
    address={Berlin, Heidelberg},
    pages={75--102},
    numpages={28},
}

@article{wu2025femtosecond,
  title={Femtosecond laser--assisted printing of hard magnetic microrobots for swimming upstream in subcentimeter-per-second blood flow},
  author={Wu, Dong and Wang, Xiuwen and Li, Rui and Wang, Chaowei and Ren, Zhongguo and Pan, Deng and Ren, Pinliang and Hu, Yanlei and Xin, Chen and Zhang, Li},
  journal={Science Advances},
  volume={11},
  number={27},
  pages={eadw1272},
  year={2025},
  publisher={American Association for the Advancement of Science}
}

@article{ding2016microfluidic,
  title={Microfluidic-based droplet and cell manipulations using artificial bacterial flagella},
  author={Ding, Yun and Qiu, Famin and Casadevall i Solvas, Xavier and Chiu, Flora Wing Yin and Nelson, Bradley J and DeMello, Andrew},
  journal={Micromachines},
  volume={7},
  number={2},
  pages={25},
  year={2016},
  publisher={MDPI}
}

@article{dulinska2019superparamagnetic,
  title={Superparamagnetic iron oxide nanoparticles—current and prospective medical applications},
  author={Duli{\'n}ska-Litewka, Joanna and {\L}azarczyk, Agnieszka and Ha{\l}ubiec, Przemys{\l}aw and Szafra{\'n}ski, Oskar and Karnas, Karolina and Karewicz, Anna},
  journal={Materials},
  volume={12},
  number={4},
  pages={617},
  year={2019},
  publisher={MDPI}
}

@article{eom2025fast,
  title={Fast 3{D} printing of fine, continuous, and soft fibers via embedded solvent exchange},
  author={Eom, Wonsik and Hossain, Mohammad Tanver and Parasramka, Vidush and Kim, Jeongmin and Siu, Ryan WY and Sanders, Kate A and Piorkowski, Dakota and Lowe, Andrew and Koh, Hyun Gi and De Volder, Michael FL and others},
  journal={Nature Communications},
  volume={16},
  number={1},
  pages={842},
  year={2025},
  publisher={Nature Publishing Group UK London}
}

@article{iwaki2016programmable,
  title={A programmable {DNA} origami nanospring that reveals force-induced adjacent binding of myosin {VI} heads},
  author={Iwaki, M and Wickham, SF and Ikezaki, K and Yanagida, T and Shih, WM},
  journal={Nature Communications},
  volume={7},
  number={1},
  pages={13715},
  year={2016},
  publisher={Nature Publishing Group UK London}
}

@article{golestanian2008analytic,
  title={Analytic results for the three-sphere swimmer at low {R}eynolds number},
  author={Golestanian, Ramin and Ajdari, Armand},
  journal={Physical Review E},
  volume={77},
  number={3},
  pages={036308},
  year={2008},
  publisher={APS}
}

@article{alexander2009hydrodynamics,
  title={Hydrodynamics of linked sphere model swimmers},
  author={Alexander, GP and Pooley, CM and Yeomans, JM},
  journal={Journal of Physics: Condensed Matter},
  volume={21},
  number={20},
  pages={204108},
  year={2009},
  publisher={IOP Publishing}
}

@article{pande2015forces,
  title={Forces and shapes as determinants of micro-swimming: effect on synchronisation and the utilisation of drag},
  author={Pande, Jayant and Smith, Ana-Sun{\v{c}}ana},
  journal={Soft Matter},
  volume={11},
  number={12},
  pages={2364--2371},
  year={2015},
  publisher={Royal Society of Chemistry}
}

@article{holm2005structure,
  title={The structure of ferrofluids: A status report},
  author={Holm, Christian and Weis, J-J},
  journal={Current Opinion in Colloid \& Interface Science},
  volume={10},
  number={3-4},
  pages={133--140},
  year={2005},
  publisher={Elsevier}
}

@article{sanchez2015supramolecular,
  title={Supramolecular magnetic brushes: the impact of dipolar interactions on the equilibrium structure},
  author={S{\'a}nchez, Pedro A and Pyanzina, Elena S and Novak, Ekaterina V and Cerd{\`a}, Joan J and Sintes, Tomas and Kantorovich, Sofia S},
  journal={Macromolecules},
  volume={48},
  number={20},
  pages={7658--7669},
  year={2015},
  publisher={ACS Publications}
}

@article{klapp2016collective,
  title={Collective dynamics of dipolar and multipolar colloids: from passive to active systems},
  author={Klapp, Sabine HL},
  journal={Current Opinion in Colloid \& Interface Science},
  volume={21},
  pages={76--85},
  year={2016},
  publisher={Elsevier}
}

@article{box2017motion,
  title={On the motion of linked spheres in a Stokes flow},
  author={Box, F and Han, E and Tipton, CR and Mullin, T},
  journal={Experiments in Fluids},
  volume={58},
  number={4},
  pages={29},
  year={2017},
  publisher={Springer}
}

@article{dreyfus2005microscopic,
  title={Microscopic artificial swimmers},
  author={Dreyfus, R{\'e}mi and Baudry, Jean and Roper, Marcus L and Fermigier, Marc and Stone, Howard A and Bibette, J{\'e}r{\^o}me},
  journal={Nature},
  volume={437},
  number={7060},
  pages={862--865},
  year={2005},
  publisher={Nature Publishing Group UK London}
}

@article{goubault2003flexible,
  title={Flexible magnetic filaments as micromechanical sensors},
  author={Goubault, C{\'e}cile and Jop, Pierre and Fermigier, Marc and Baudry, Jean and Bertrand, Emanuel and Bibette, J{\'e}r{\^o}me},
  journal={Physical review letters},
  volume={91},
  number={26},
  pages={260802},
  year={2003},
  publisher={APS}
}

@article{amoudruz2022independent,
  title={Independent control and path planning of microswimmers with a uniform magnetic field},
  author={Amoudruz, Lucas and Koumoutsakos, Petros},
  journal={Advanced Intelligent Systems},
  volume={4},
  number={3},
  pages={2100183},
  year={2022},
  publisher={Wiley Online Library}
}




\end{document}